\documentstyle[twocolumn,prc,aps,psfig]{revtex}
\begin{document}
\title{Probing the antikaon potential by K$^-$A elastic
scattering\thanks{Supported by Forschungszentrum J\"ulich}}
\author{A. Sibirtsev  and W. Cassing}
\address{Institute for Theoretical Physics, \\
University of Giessen, D-35392 Giessen, Germany}
\date{\today}
\maketitle

\begin{abstract}
We extract the antikaon potential from differential data on 
elastic $K^-A$ scattering within  Glauber theory at an antikaon 
momentum of 800~MeV/c.  The $K^-N$ cross section at densities up 
to 0.1 $fm^3$ is found to be close
to the value given in free space $K^-p$ and $K^-n$ interactions 
(averaged over the number of protons and neutrons) whereas
the ratio $\alpha$ of the real to imaginary part of the forward
scattering amplitude $f(0)$ differs from the ratio $\alpha$ in free space
indicating a dropping of $Re f(0)$ at finite nuclear density.
The extrapolated $K^-$ potential at a momentum of 0.8 GeV/c amounts to
$\approx $--28 MeV at density $\rho_0$ which is substantially less than
the values extracted from kaonic atoms at zero momentum or heavy-ion data
at 300 MeV/c${\leq}p_K{\leq}$600 MeV/c.
We, furthermore, investigate the perspectives of scattering experiments 
on $^{12}C$ at low ($\approx$300 MeV/c) and high ($\approx$1.2 GeV/c) 
momenta.
\end{abstract}

\pacs{ PACS: 11.55.Fv; 13.75.Jz; 13.85.Dz; 25.80.-e \\
Keywords:  dispersion relations; kaon-baryon interactions; 
Elastic scattering; meson-induced reactions}

\section{Introduction}
Since the pioneering work of Kaplan and Nelson \cite{Kaplan}, 
that predicted large attractive antikaon self-energies in nuclear 
matter, speculations about strange condensates in neutron stars and 
relativistic heavy-ion collisions have raised the interest in 
kaon-nuclear physics~\cite{Brown1,Brown2,Thorsson,Schaffner}. 
These ideas have been 
supported by the data on kaonic atoms that indicate a strong attractive
potential at zero antikaon momentum~\cite{Gal1,Gal2}. On the other 
hand, also the data on $K^-$ production from heavy-ion 
collisions~\cite{Schroter,Senger1,Barth,Laue,Senger2} 
have been interpreted in terms of a dropping antikaon mass at finite 
density~\cite{Li1,Cassing1,Li2,Bratkovskaya,Cassing2}. 
Whereas the analysis~\cite{Gal1,Gal2} of the kaonic 
atom data suggests an attractive antikaon potential of 
$\simeq$--180~MeV at normal nuclear density,  the data 
on $K^-$  production from heavy-ion collisions lead to 
a  potential  $\simeq$--80$\div$120~MeV
\cite{Li1,Cassing1,Li2,Bratkovskaya,Cassing2}.
This discrepancy might be attributed~\cite{Sibirtsev1,Sibirtsev4} 
to the momentum dependence of the antikaon potential, since the kaonic
atoms explore stopped antikaons with $p_K{\approx}$0, while the heavy-ion
experiments have probed the range 300${\le}p_K{\le}$600~MeV/c.
 
Among the available experimental methods the elastic and
inelastic hadron-nucleus scattering are a traditional 
way~\cite{Glauber1} to study the optical potential.
Moreover, by measuring the $hA$ scattering at different
projectile momenta one can also evaluate the momentum-dependence 
of the potential. Here we study the possibility
to measure the $K^-$ optical potential by elastic $K^-A$
scattering,  analyze the available data and discuss the
perspectives of $K^-$-meson scattering experiments at low
and high momenta.

\section{Glauber approximation}
The elastic scattering of an energetic hadron from a nucleus can 
be described by the Glauber theory of diffractive scattering as
formulated in the Boulder Lectures in Theoretical 
Physics~\cite{Glauber1}. Here we refer to the final results from 
Ref.~\cite{Glauber1} and present the formulas relevant 
for the further calculations and  discussions. 

When neglecting the Coulomb distortion of the projectile the amplitude 
for the elastic scattering of the particle from the nucleus is given as
\begin{equation}
F_{el}(q) = ik \int\limits_0^\infty J_0(bq) \,
\left(1-\exp[i\,\chi_N(b)\,]\,\right)\, b \, db,
\label{el1}
\end{equation} 
where $q$ is the momentum transfered from the 
projectile to the nucleus, $k$ is the missile momentum, 
while $J_0$ is the zero'th order
Bessel function. The phase shift $\chi_N$ may be
approximated by
\begin{equation}
\chi_N(b)=\frac{2 \pi \, f(0)}{k}  \int\limits_{-\infty}^\infty
\rho(b,z) \,dz,
\label{shift1}
\end{equation}
where $f(0)$ is the complex amplitude for the forward scattering 
of the projectile on a proton or neutron in the nucleus, while
$\rho(r{=}\sqrt{b^2{+}z^2})$ is  the nuclear density distribution.
In principle,  $f(0)$ should be given by the (density dependent) in-medium
amplitude, but within the impulse approximation~\cite{Newton} 
it can be approximated by the forward scattering amplitude 
in free space. Via the optical theorem, furthermore, the imaginary part 
of $f(0)$ can be expressed in terms of the  total particle-nucleon
cross section $\sigma$ as  
\begin{equation}
Im f(0) = \frac{k}{4\pi} \, \sigma.
\end{equation} 
Introducing $\alpha{=}Ref(0)/Imf(0)$ the phase~(\ref{shift1})
might be rewritten as
\begin{equation}
\chi_N(b) = \sigma \ \frac{\alpha +i}{2} 
\int\limits_{-\infty}^\infty
\rho(b,z) \,dz.
\label{shift2}
\end{equation}

The experimental results on the elastic differential cross section
given by
\begin{equation}
\frac{d\sigma}{d\Omega} = |F_{el}(q)|^2
\label{dif}
\end{equation}
(with $q{=}2k \sin(\theta/2)$) can be used for the evaluation
of the cross section $\sigma$ and the ratio of the real to imaginary
forward scattering amplitude $\alpha$. Recall, that $\sigma$ and 
$\alpha$ might differ from the relevant values given in free space; 
a question that has to be discussed separately.
Neglecting such modifications for a while, the traditional 
Glauber approximation thus allows to evaluate the 
in-medium parameters describing the interaction of the
projectile with a finite nucleus.

Furthermore, the Coulomb correction may be taken into account 
by adding  the Coulomb phase shift to the nuclear phase as  
proposed in Refs.~\cite{Czyz,Glauber2,Ahmad,Alkhasov,Dalkarov1}. 
The Coulomb corrected nuclear elastic scattering amplitude 
then is given by
\begin{eqnarray}
F_{el}(q)=F_C(q)+ik\int\limits_0^\infty J_0(bq) \, 
\exp[i\chi_C(b)\,] \nonumber \\ \times 
\left(1{-}\exp[i\chi_N(b){+}i\chi(b)\,]\,\right)\,b \,db.
\label{el2}
\end{eqnarray}
The  amplitude $F_C$ and the phase shift $\chi_C$ describe
the scattering by a point charge $Z$,
\begin{equation}
F_C(q){=}{-}\frac{2\xi k}{q^2}\,\exp[i\phi_C], \hspace{11mm}
\chi_C(b){=}2\,\xi\, ln (kb),
\end{equation}
where 
\begin{equation}
\phi_C{=}{-}2\,\xi\,ln\frac{q}{2k}{+}2\eta, \hspace{3mm}
\xi{=}{-}\frac{Ze^2m}{k\,\hbar c}, \hspace{3mm} 
\eta{=}arg\Gamma(1{+}i\xi).
\end{equation}
The sign of $\xi$ is given by the attractive (negative)
or repulsive (positive) interaction of the projectile with the
Coulomb field. 

The phase shift $\chi$ in Eq.~(\ref{el2}) is given by
\begin{equation}
\chi{=}\frac{8\pi \xi}{A}\int\limits_b^\infty\!\rho(r)\,
\left(ln\left[\frac{1{+}\sqrt{1{-}\tau^2}}{\tau}\right]{-}
\sqrt{1-\tau^2}\right)
r^2\,dr,
\label{shift4}
\end{equation}
with $\tau{=}b/r$ while $\rho(r)$ is the charge distribution. 

We assume the charge density in Eq.~(\ref{shift4}) to
be proportional to the nuclear density function in Eq.~(\ref{shift1}),
which is normalized to the total number of 
nucleons $A$. For 4${\le}A{\le}$16 nuclei - to a good approximation - 
it is given by~\cite{Dalkarov1}
\begin{equation}
\rho(r)=\left(R\,\sqrt{\pi}\right)^{-3}\left[4+
\frac{2\,(A-4)\,r^2}{3\,R^2}\right] 
\exp{\left[-r^2/R^2\right]},
\label{light}
\end{equation}
with $R{=}\sqrt{2.5}$~fm for $^{12}C$. For $A{>}$16 nuclei the
nuclear density is replaced by the Wood-Saxon distribution
\begin{equation}
\rho(r)=\frac{\rho_0}{1+\exp[(r-R)/d]}
\label{heavy}
\end{equation}
with
\begin{equation}
R{=}1.28A^{1/3}{-}0.76{+}0.8A^{-1/3}\ \mbox{fm}, \hspace{5mm}
d{=}\sqrt{3}/\pi\ \mbox{fm}.
\end{equation}               

We note that we do not account for a screening phase and discard 
nucleus recoil corrections~\cite{Alkhasov,Dalkarov1}.

\section{Data analysis}
\begin{figure}[h]
\vspace{-4mm}
\psfig{file=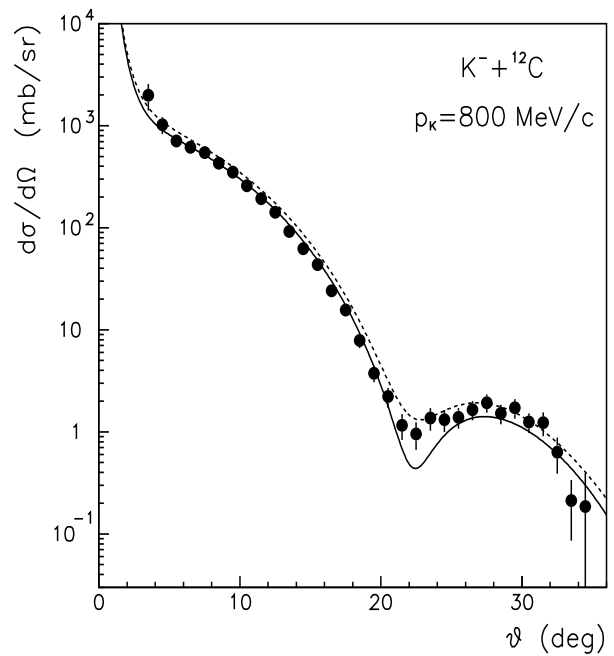,width=9cm,height=10.cm}
\vspace{-3mm}
\caption[]{\label{elastic1}The elastic differential cross section
for $K^-$ scattering from $^{12}C$. The data are from 
Ref.~\protect\cite{Marlow}. The solid line shows the
result from the fit with the parameters listed in 
Table.~\protect\ref{tab1} while the dashed line indicates the
calculations with $\sigma$ and $\alpha$ taken from 
$K^-N$ scattering in free space averaged over protons and neutrons in the
target.}
\vspace{2mm}
\end{figure}

\begin{figure}[h]
\vspace{-5mm}
\psfig{file=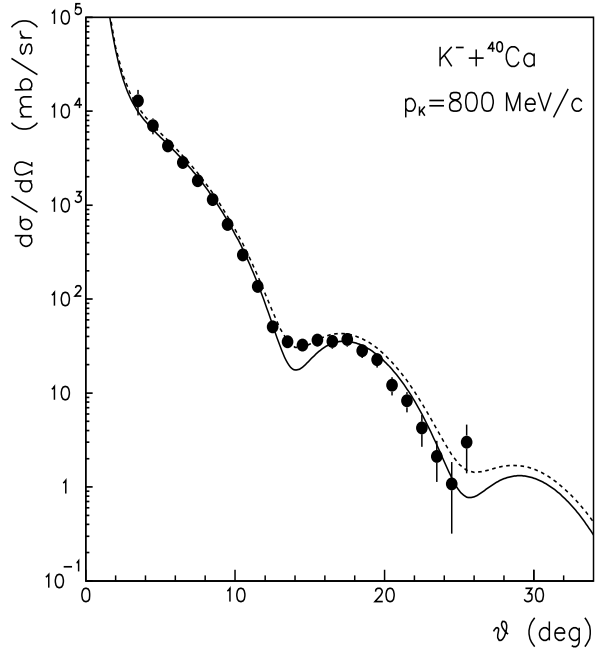,width=9cm,height=10.cm}
\vspace{-3mm}
\caption[]{\label{elastic2}The differential cross section
for $K^-$ elastic scattering from $^{40}Ca$. The notations
are the same as in Fig.~\protect\ref{elastic1}.}
\end{figure}

The data on the differential cross section
for the $K^-$-meson elastic scattering by $^{12}C$
and $^{40}Ca$ at an antikaon momentum of 800~MeV/c have been
collected by Marlow et al.~\cite{Marlow} and
fitted by Eq.~(\ref{el2}) in order to evaluate the
cross section $\sigma$ and real to imaginary ratio $\alpha$.
We here use the Minuit~\cite{James} procedure and the $\chi^2$ 
method. The solid lines in Figs.~\ref{elastic1},\ref{elastic2} 
show the result from the minimization procedure together with
the experimental results~\cite{Marlow}. The parameters 
$\sigma$ and $\alpha$ as well as the reduced  $\chi^2$ are listed 
in Table~\ref{tab1}. 

The dashed lines in  Figs.~\ref{elastic1},\ref{elastic2} 
show the calculations with $\sigma$ and $\alpha$ taken from the
$K^-p$ and $K^-n$ interactions in free space and averaged 
over the number of  protons and neutrons in the target as 
described below. We note, that although the 
calculations with the free space variables provide a better 
description in the vicinity of the diffractive minima, they 
lead to a substantially worse $\chi^2/N$ as shown in Table~\ref{tab1}.

One of the  crucial questions now is the sensitivity of the
data to the sign and the magnitude of the ratio $\alpha$.
It is known~\cite{Glauber1} that - while $\sigma$ is
given by the absolute value of the nuclear cross section - the ratio
$\alpha$ may be fixed by the differential cross section
at the diffractive minima. Actually, any effect providing
a complex correction of the nuclear scattering
amplitude $F_{el}$ influences the filling of the
diffractive minima~\cite{Dalkarov1,Dalkarov2}.

\begin{figure}[h]
\vspace{-4mm}
\psfig{file=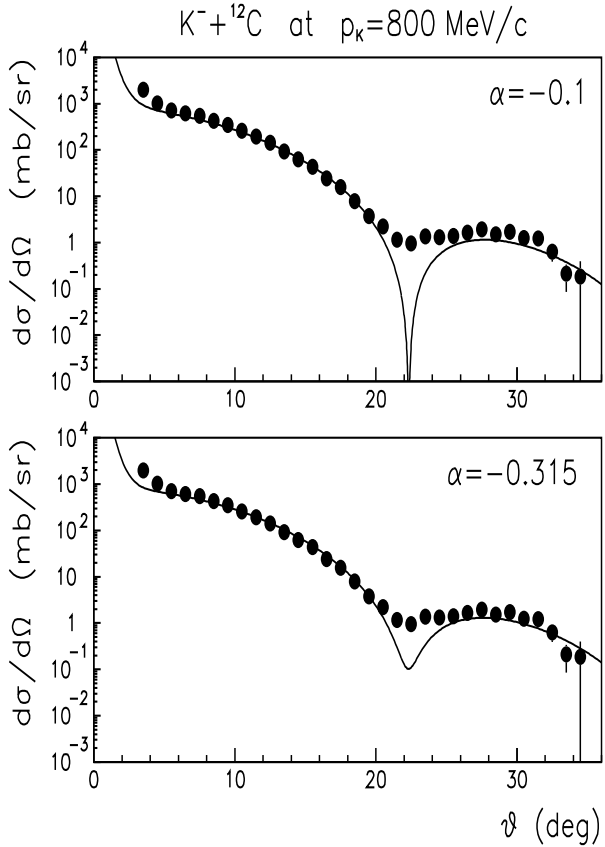,width=9cm,height=12.5cm}
\vspace{-3mm}
\caption[]{\label{elastic1a}The differential cross section
for  $K^-$ elastic scattering from $^{12}C$. The
data are from Ref~\protect\cite{Marlow}. The lines show 
calculations with $\alpha$=-0.1 (upper part) and $\alpha$=-0.315
(lower part).}
\end{figure}

Fig.~\ref{elastic1a} shows the $K^-C$ elastic differential
cross section calculated for ratios $\alpha$=-0.1 and
$\alpha$=-0.315. The latter value for $\alpha$ was taken to be equal 
in magnitude 
to the parameter evaluated by the minimization procedure, but 
by substituting an opposite  sign in order to demonstrate the 
sensitivity of the data to the sign of $\alpha$. 

\begin{table}[h]
\caption{\label{tab1} The $K^-N$ total cross section $\sigma$,
ratio $\alpha$ of the real to imaginary part of the forward $K^-N$
scattering amplitude and $\chi^2/N$ evaluated from the data on $K^-A$
elastic scattering and that taken from the interaction in free space.}
\vspace{2mm}
\begin{tabular}{lcccc}
& \multicolumn{2}{c}{evaluated from}&
\multicolumn{2}{c}{taken from free} \\
\vspace{2mm}
& \multicolumn{2}{c}{$K^-A$ data~\protect\cite{Marlow}} & 
\multicolumn{2}{c}{$K^-N$ interactions} \\
\vspace{2mm}
& $^{12}C$ & $^{40}Ca$ & $^{12}C$ & \hspace{4mm} $^{40}Ca$ \\
\vspace{2mm}
$\alpha$ & 0.315$\pm$0.001 & 0.396$\pm$0.003 &
\multicolumn{2}{c}{0.608} \\
\vspace{2mm}
$\sigma$ & 37.97$\pm$0.01 & 38.3$\pm$0.2 &
\multicolumn{2}{c}{36.12} \\
\vspace{2mm}
$\chi^2/N$ & 1.4 & 1.7 & 5.1 & \hspace{4mm}4.3 
\end{tabular}
\end{table}

Now Fig.~\ref{elastic1a} illustrates that the absolute 
value of the ratio $\alpha$ might be reasonably fixed by the data.
The calculations with $\alpha$=-0.315 provide a $\chi^2/N$=2.3,
which is worse as compared to the value listed in Table~\ref{tab1}.
Indeed, the data are sensitive to both sign and magnitude
of the ratio $\alpha$ when taking into account the Coulomb
scattering. This method has been
used before for the evaluation of the real to imaginary ratio 
$\alpha$ in case of the antiproton-nucleon forward scattering 
amplitude in Refs.~\cite{Dalkarov1,Dalkarov2,Cresti,LEAR,Iwasaki}.

\begin{figure}[h]
\vspace{-10mm}
\psfig{file=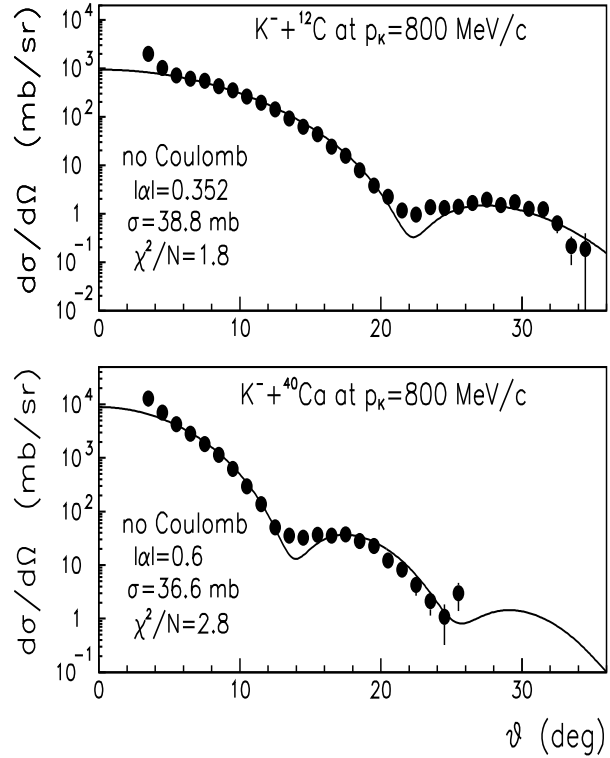,width=9cm,height=12.5cm}
\vspace{-3mm}
\caption[]{\label{elastic1b}The differential cross section
for $K^-$ elastic scattering from $^{12}C$ and 
$^{40}Ca$. The data are from Ref~\protect\cite{Marlow}. The lines 
show calculations without the Coulomb correction. The parameters
$\sigma$ and $\alpha$  evaluated by  
Minuit~\protect\cite{James} are shown in the figure
together with the respective $\chi^2/N$.}
\end{figure}

Since the inclusion of the Coulomb 
correction is important for the determination of the sign
of $\alpha$, we also have performed calculations taking 
into account only the nuclear phase shift.
The results of this limit are shown in Fig.~\ref{elastic1b} for $K^-$
scattering from $^{12}C$ and $^{40}Ca$ together with
the data~\cite{Marlow}. The parameters $\sigma$ and $\alpha$
(extracted from the data) as well as the associated $\chi^2/N$ are also 
shown in Fig.~\ref{elastic1b}. Note, that when neglecting the
Coulomb correction the nuclear data  become insensitive
to the sign of $\alpha$; furthermore, for
$\alpha$=0 the differential cross section in 
the diffractive minima will be zero.

  From the analysis presented above it is clear, that without the 
Coulomb scattering it is not possible to reproduce the 
data at low $q$. Although there are not much data available at 
small scattering angles, the Coulomb effect by obvious 
reason must be included in the data analysis.  We also note
that the calculations with the Coulomb 
effect yield the minimal $\chi^2/N$.

We conclude that the data~\cite{Marlow} on elastic $K^-$-meson
scattering from $^{12}C$ and $^{40}Ca$ nuclei are sufficiently
sensitive to the evaluation of the in-medium total $K^-N$
cross section $\sigma$ and the ratio $\alpha$ of the real
to imaginary part of the in-medium forward scattering amplitude $f(0)$.
The resulting parameters are averaged over the proton and neutron
numbers of the target. In the following these in-medium
parameters will be compared to those extracted from
the $K^-N$ interaction in  free space.

\section{What densities can be probed by K$^-$A scattering?}
An important question in $K^-A$ scattering is to what extent one 
can probe the nuclear interior. Obviously, for very peripheral 
scattering the values for $\sigma$ and $\alpha$ should be 
sufficiently close to those values  given by the $K^-N$ interaction 
in free space.

\begin{figure}[h]
\vspace{-2mm}
\begin{minipage}[h]{9cm}
\hspace{-6mm}\psfig{file=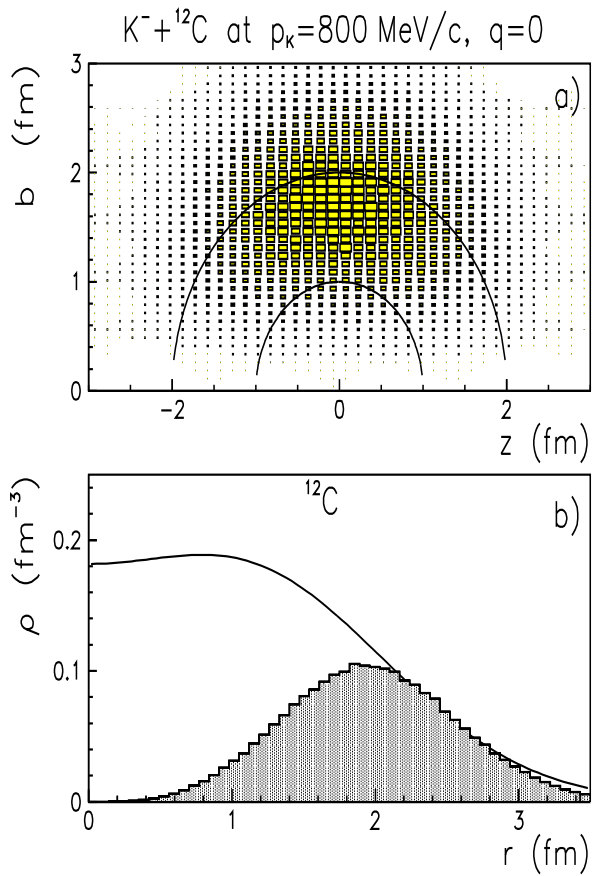,width=9.2cm,height=13.1cm}
\end{minipage}
\vspace{-5mm}
\caption[]{\label{elastic7}a) The differential cross section
for $K^-$ elastic scattering from $^{12}C$ at an antikaon
momentum $p_K$=800~MeV/c and $q$=0 as a function of the
impact parameter $b$ and coordinate $z$ along the beam axis.
The sizes of the granulated boxes are proportional to the
cross section; the solid lines indicate
radii of 1 and 2~fm.
b) The solid line shows the density 
profile for $^{12}C$ while the histogram displays the scattering 
profile normalized to the density tail at large $r$.}
\end{figure}

Indeed, Eq.~(\ref{el1}) indicates that elastic scattering is a 
peripheral process~\cite{Kondratyuk}. This is illustrated by
Fig.~\ref{elastic7}a), where the differential cross section for
$K^-$-meson scattering from $^{12}C$ at $\theta$=0 for an antikaon 
momentum of 800~MeV/c  is shown as a function of the impact 
parameter $b$ and the coordinate $z$ along the beam axis 
(see Eq.~(\ref{el1})). The calculations show that the scattering 
is predominantly peripheral. To obtain
the density probed by  $K^-$ scattering at this momentum we
show (solid line in Fig.~\ref{elastic7}b)) the 
density profile for $^{12}C$ given by Eq.~(\ref{light}).
The hatched histogram in Fig.~\ref{elastic7}b)
indicates the $K^-C$ scattering profile distribution, which 
was calculated by taking the $K^-C$ elastic 
differential cross section at $p_K$=800~MeV/c and $q$=0 at 
given coordinates $b$ and $z$ and by integrating over the contour 
$r{=}\sqrt{b^2{+}z^2}$. The scattering profile is normalized to 
the tail of the density function $\rho$.

\begin{figure}[h]
\vspace{-3mm}
\begin{minipage}[h]{9cm}
\hspace{-6mm}\psfig{file=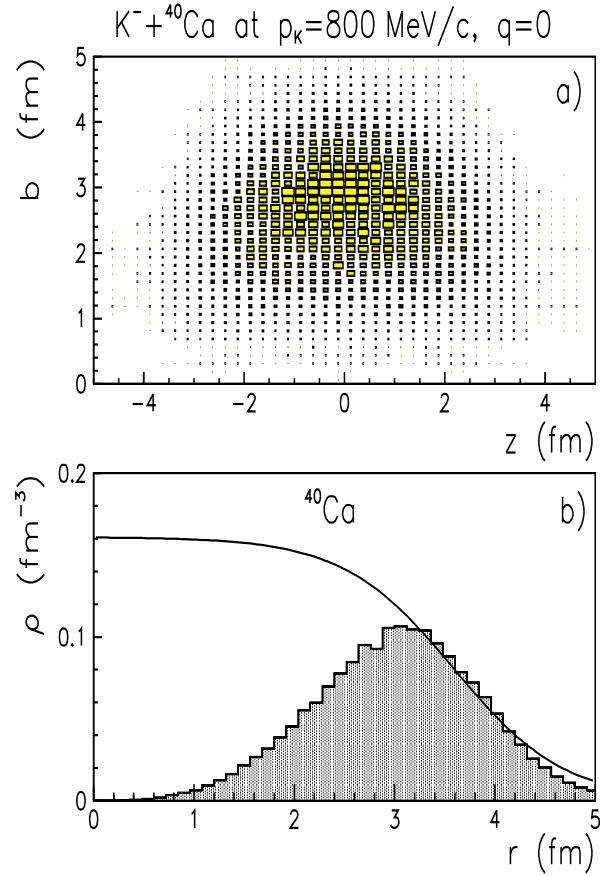,width=9.2cm,height=13.2cm}
\end{minipage}
\vspace{-7mm}
\caption[]{\label{elastic7a}Same as in 
Fig.~\protect\ref{elastic7} for $K^-$ scattering from
$^{40}Ca$.}
\end{figure}

The maximal nuclear density that might be tested by 
$K^-$ elastic scattering from $^{12}C$ is about
$\simeq$0.1~fm$^{-3}$, that should be compared to the saturation  
value $\rho_0$=0.16~fm$^{-3}$. The average densities
probed by the $K^-$ scattering are in the range of  $\rho_0$/3.
Similar conclusions also hold for the $^{40}Ca$ nucleus. 
This is demonstrated in Fig.~\ref{elastic7a} which shows 
the calculations of the reaction 
zone and the profile functions for $K^-$ scattering from $^{40}Ca$.

Now the in-medium modification of the total cross section
$\sigma$ and the ratio $\alpha$ of the real to imaginary part 
of the forward scattering amplitude and their difference from 
those values in free space might be discussed for
the baryon densities probed in $K^-A$ scattering.

\section{The K$^-$N forward scattering amplitude in free space}
The forward $K^-$ scattering amplitude on the proton and the neutron
has been calculated in our previous study~\cite{Sibirtsev1}. 
Here, we again evaluate the imaginary part of $f(0)$  from the
available experimental data on the total cross section, while 
$Ref(0)$ is calculated by the dispersion relations. The
details of the calculations and references to related dispersion 
analyses are given in Ref.~\cite{Sibirtsev1}. We directly 
continue with the status of the data and proper parameterizations.

\begin{figure}[h]
\vspace{-6mm}
\psfig{file=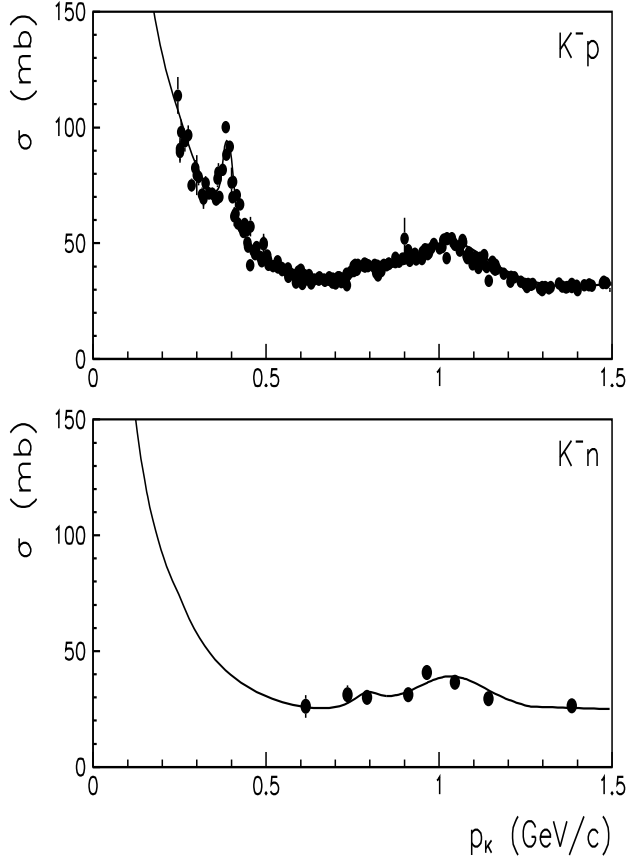,width=9.2cm,height=13.4cm}
\vspace{-4mm}
\caption[]{\label{elastic8}The data on $K^-p$ and $K^-n$ total 
cross sections~\protect\cite{PDG} together with the
parameterizations from Ref.~\protect\cite{Sibirtsev1}.}
\end{figure}

Fig.~\ref{elastic8} shows the experimental results~\cite{PDG} on the
$K^-p$ and $K^-n$ total cross sections together with the
parameterizations from Ref.~\cite{Sibirtsev1}. Notice, that
the data are well defined at $p_K$=800~MeV/c.

\begin{figure}[h]
\vspace{-6mm}
\psfig{file=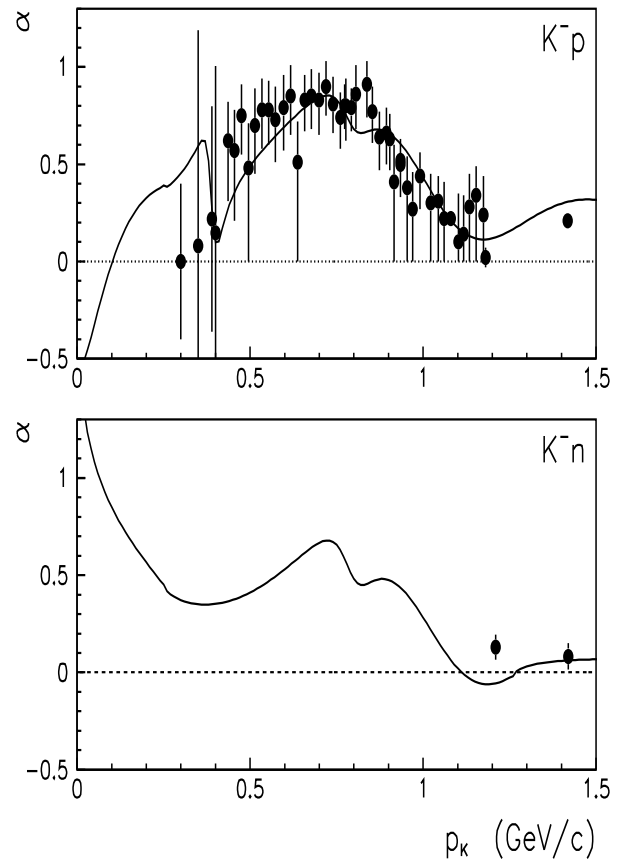,width=9.2cm,height=13.5cm}
\vspace{-5mm}
\caption[]{\label{elastic8a}The ratio $\alpha$ of the real to 
imaginary part for the  $K^-p$ and $K^-n$  forward scattering 
amplitudes. The experimental results are taken from 
Refs.~\protect\cite{Conforto,Armenteros,Dumbrais,Martin}
while the solid lines show our calculations~\protect\cite{Sibirtsev1}.}
\end{figure}

The resulting ratio $\alpha{=}Ref(0)/Imf(0)$ is shown in 
Fig.~\ref{elastic8a}
as a function of the antikaon momentum for a proton and
neutron target; the  $K^-p$  experimental results 
are taken from Refs.~\cite{Conforto,Armenteros,Dumbrais}.

It is important to note, that the experimental result on the
real part of the forward scattering amplitude is determined from
\begin{equation}
\left[\,Re f(0)\,\right]^2 = \left. \frac{d\sigma}{d\Omega} 
\right|_{q=0} - \,\left[\frac{k \sigma}{4 \pi}\right]^2,
\end{equation}
which does not provide the sign of $\alpha$. However,
the sign of $Ref(0)$ can be fixed by the combined data analysis 
with the $K$-matrix and the dispersion relations~\cite{Martin}
by applying the crossing symmetry.
We also note, that the amount of experimental points shown in
Fig.~\ref{elastic8a} for $K^-p$ scattering is larger then
that in the latest analysis by Martin~\cite{Martin}.

The situation for the real part of the $K^-n$ forward scattering 
amplitude is different. While $Re(f(0)$ can be reasonably 
determined from the low energy solution within 
the $K$-matrix analysis, there remains
an uncertainty in the absolute normalization of the substraction
point. The latter one is fixed by the two experimental 
points~\cite{Martin} shown in the lower part of  
Fig.~\ref{elastic8a}. However, the experimental
results do not indicate the sign of $Ref(0)$. 

Furthermore,
there are no data available for the real part of the $K^+n$
forward scattering amplitude. 
In principle, the ratio $\alpha$ for the $K^-n$ scattering
might stay positive and not change its sign around 
$p_K{\simeq}$1.2~GeV. However, for these
antikaon momenta the ratio $\alpha$ is small and close to zero. 
This important feature will be discussed later again.

Keeping in mind the uncertainties in the determination of the
ratio $\alpha$ for the $K^-n$ interaction we now can perform 
a comparison with the data evaluated from the $K^-$ elastic
scattering on $^{12}C$ and $^{40}Ca$.

\section{Comparison with the data from K$^-$A scattering} 
Both, $^{12}C$ and $^{40}Ca$, are isospin-symmetric nuclei and 
for a comparison with the values of $\sigma$ and $\alpha$ 
in free space we average over the results from
the previous section for proton and neutron targets.
In principle, the cross section $\sigma$ and the ratio $\alpha$
should be averaged over the Fermi distribution available at 
densities $\rho{<}$0.1~fm$^{-3}$, which is important
for antikaon momenta where $\sigma$ or $\alpha$ 
rapidly change with $p_K$. However, this is not the 
case for $p_K$=800~MeV/c.

\begin{figure}[h]
\vspace{-6mm}
\psfig{file=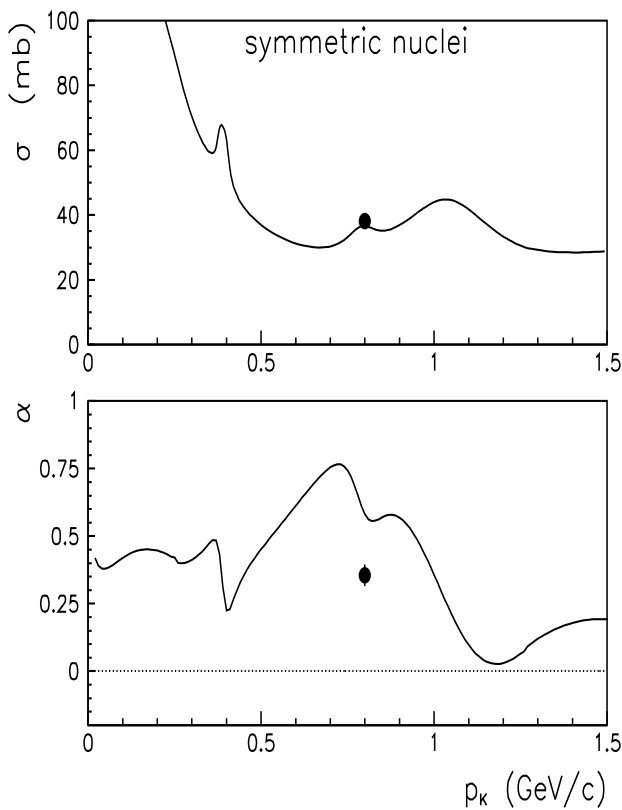,width=9.2cm,height=12.5cm}
\vspace{-4mm}
\caption[]{\label{elastic8b}The solid lines indicate the
$K^-N$ cross section $\sigma$ and ratio $\alpha$ given in free 
space  averaged over  equal numbers of protons and 
neutrons. The solid circles show the results evaluated from
the data on $K^-$ elastic scattering from nuclei
averaged over $^{12}C$ and $^{40}Ca$.}
\end{figure}

The cross section $\sigma$ and ratio $\alpha$ taken from the
free $K^-N$ interactions averaged over equal numbers 
of protons and neutrons are shown in Fig.~\ref{elastic8b} 
by the solid lines. The dots in Fig.~\ref{elastic8b} 
indicate the results evaluated from the data on $K^-$ elastic 
scattering (see Table~\ref{tab1}) averaged over the 
$^{12}C$ and $^{40}Ca$ targets.

We find that the total cross section extracted from nuclear elastic 
scattering is practically equal to the value given in free space, 
while the ratio $\alpha$ differs by about a factor of $\simeq$1.5.
When discussing this difference one should keep in mind
that the calculations actually do not reproduce the data exactly 
at the diffractive minima leaving room for an
uncertainty in the ratio $\alpha$, which is not 
well defined in free $K^-$-neutron interactions. The factor 
$\simeq$1.5 might be attributed to the substraction point providing
a global shift of $\alpha$ (cf. Fig.~\ref{elastic8a}).
However, in this case the two experimental points~\cite{Martin} 
for the real part of the $K^-n$ forward scattering 
amplitude cannot be reproduced. We thus reject this possibility.

Since the total $K^-N$ cross section apparently is not modified 
in nuclear matter at densities $\rho{<}$0.1~fm$^{-3}$ and
antikaon momenta of 800~MeV/c relative to its
value in free space, the drop in $\alpha$ should be attributed 
to a decrease of the
real part of the forward $K^-N$ scattering amplitude.

\section{The K$^-$ optical potential}
Adopting a spherically symmetric nuclear optical potential $U_N$ 
the phase shift $\chi_N$ is given as~\cite{Glauber1}
\begin{equation}
\chi_N(b)= -\frac{m}{k}\int\limits_{-\infty}^\infty
U_N(b,z) \,dz.
\label{pot}
\end{equation}
Comparing to Eq.~(\ref{shift1}) this gives
\begin{equation}
U_N(r)=-\frac{2\pi f(0)}{m} \, \rho(r),
\end{equation}
which can be rewritten in terms of $\sigma$ and $\alpha$ as
\begin{equation}
U_N(r)=-k\,\sigma \, \frac{\alpha + i}{2m} \rho(r).
\end{equation}
The Coulomb potential, furthermore,  is given as~\cite{Dalkarov1}
\begin{equation}
U_C(r)=-4\pi Ze^2\left[\frac{1}{r}\int\limits_0^r\!\rho(x)\,x^2\,dx+
\int\limits_r^\infty\!\rho(x)\,x\,dx\right],
\end{equation}
where $\rho$ is the density distribution. 

\begin{figure}[h]
\phantom{aa}\vspace{-10mm}
\psfig{file=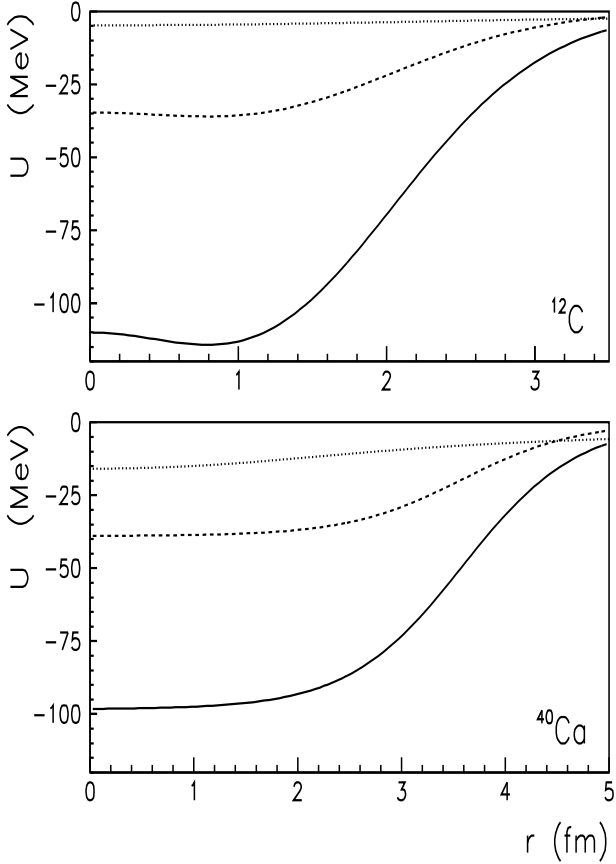,width=9.2cm,height=13.5cm}
\vspace{-4mm}
\caption[]{\label{elastic9}The imaginary (solid) and real
(dashed) nuclear potential and Coulomb potential (dotted line)
for $K^-$ scattering on $^{12}C$ and $^{40}Ca$ at an
antikaon momentum of 800~MeV/c.}
\end{figure}

Fig.~\ref{elastic9} shows the resulting real and imaginary part of the
nuclear potential and the Coulumb potential for $K^-$-meson
scattering on $^{12}C$ and $^{40}Ca$ at an antikaon momentum
of 800~MeV/c. The results for $U_N$ are shown for 
$\sigma$ and $\alpha$ as evaluated from the experimental data
on the differential elastic cross section and given in Table~\ref{tab1}.

The Coulomb potential is very small, however, 
important for the description of the differential data on
the $K^-$ elastic scattering from nuclei.
The imaginary part of the optical potential clearly dominates and is 
close to that expected from free-space calculations.
The real part of the potential is attractive but smaller than 
expected from  free-space calculations~\cite{Sibirtsev1}. 

\section{Further perspectives}
In this Section we  
outline further perspectives of experiments on 
differential data for  $K^-$ elastic scattering from nuclei. 
There are two regions of 
antikaon momenta that are crucial for the understanding of the
$K^-$ properties in nuclear matter and which can be studied by 
$K^-$-meson elastic scatterings from nuclei at densities
$\rho{<}$0.1~fm$^{-3}$. Let us discuss both of them by 
starting with the high energy region.

\subsection{Antikaon momenta around 1.2~GeV/c}
As shown in Fig.~\ref{elastic8a} the ratio $\alpha$ of
the real to imaginary part of the forward scattering amplitude
in free $K^-$-neutron interactions is close to zero for 
antikaon momenta of $\simeq$1.2~GeV/c. After  averaging of  
$\alpha$ over the number of  protons and neutrons in the 
target  the average $\alpha \approx$ 0 remains. The result
for symmetric nuclei is shown in Fig.~\ref{elastic8b}.

We recall that 
neglecting Coulomb scattering the calculations with
$\alpha$=0 yield a differential elastic cross section which is
exactly equal to zero at the diffractive minima. Taking into 
account the interference between the nuclear and Coulomb
phase shifts the $\alpha$-dependence of the cross section in the
vicinity of the diffractive minima becomes more 
complicated~\cite{Dalkarov1}, but can be predicted using
$\sigma$ and $\alpha$ given in free space. 

\begin{figure}[h]
\vspace{-4mm}
\psfig{file=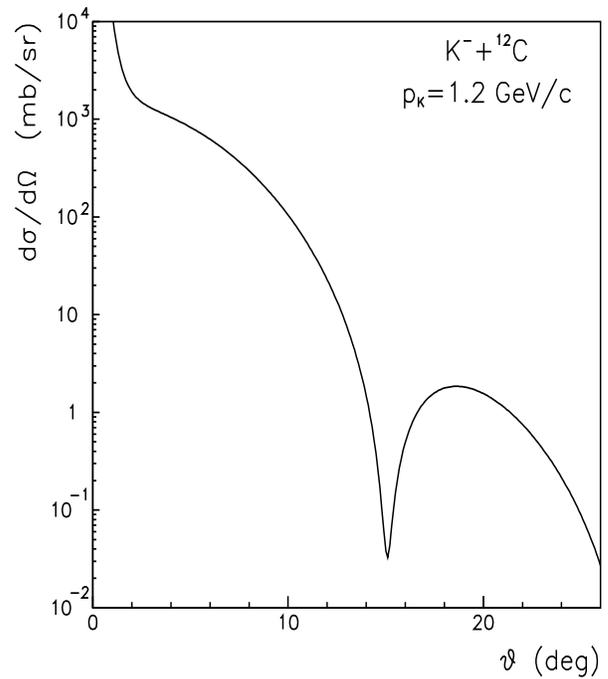,width=9cm,height=10.4cm}
\vspace{-3mm}
\caption[]{\label{elastic2a}The predicted differential cross section
for $K^-$ elastic scattering from $^{12}C$ at an antikaon
momentum of 1.2~GeV/c.}
\end{figure}

Fig.~\ref{elastic2a} shows the differential elastic cross section
for $K^-$-meson scattering from $^{12}C$ at $p_K$=1.2~GeV/c
calculated with $\sigma$=33.8~mb and $\alpha$=0.02, which corresponds to
$K^-N$ interactions in free space averaged over the number
of protons and neutrons. 

The calculations indicate a deep 
diffractive minimum due to the small value of the
ratio $\alpha$. We do expect that the measurements should
allow to extract the actual value of $\alpha$ and 
its sign. This is important for the in-medium modification 
of the real part of
the forward scattering amplitude $f(0)$ as well as for a possible
change of sign of $Ref(0)$ around
$p_K{\simeq}$1.2~GeV for the $K^-n$ interaction.

A more sophisticated analysis can be done by scanning the antikaon
momentum from 0.7 GeV/c to 1.4 GeV/c and a determination of the
differential cross section at the first diffractive minimum, i.e.
at its minimal point within the experimental resolution. As 
shown for the case of symmetric nuclear matter in Fig.~\ref{elastic8b} 
the cross section $\sigma$ does not vary very much for antikaon 
momenta in this range while the ratio $\alpha$
strongly depends on $p_K$. 

\begin{figure}[h]
\vspace{-3mm}
\psfig{file=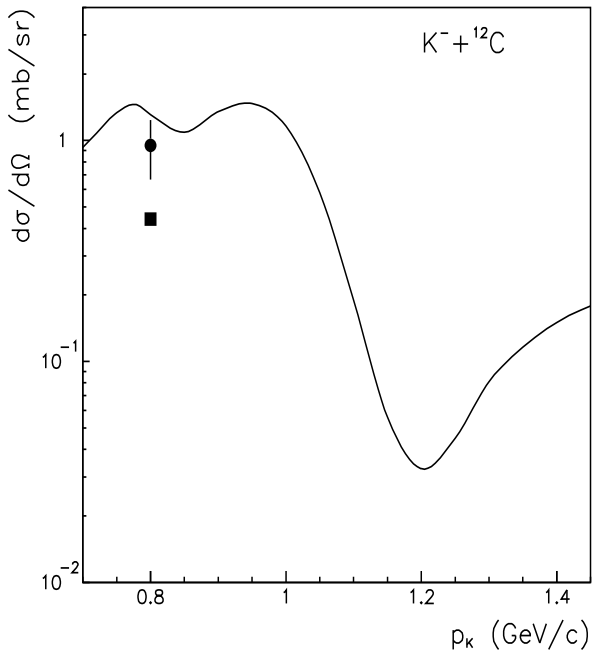,width=9cm,height=10.2cm}
\vspace{-2mm}
\caption[]{\label{elastic2b}The solid line shows the predicted 
differential cross section at the first diffractive minimum
for $K^-$ elastic scattering from $^{12}C$ as a
function of  the antikaon momentum. The calculations are performed 
with $\sigma$ and $\alpha$ given in free space. The circle shows 
the experimental result from Ref.~\protect\cite{Marlow} while
the full square shows the result from our fitting procedure at 0.8 GeV/c.}
\end{figure}

Fig.~\ref{elastic2b} shows the differential cross section at the
first diffractive minimum for $K^-$ elastic scattering from $^{12}C$ 
in the antikaon momentum range 
0.7${\le}p_K{\le}$1.4~GeV. The solid line indicates the calculations
performed with $\sigma$ and $\alpha$ taken from the
free space $K^-N$ scatterings. The calculations are averaged 
at the diffractive minimum over the scattering angles with 
$\Delta\theta{=}{\pm}$0.05 degrees. The full dot in Fig.~\ref{elastic2b}
shows the experimental result from Ref.~\cite{Marlow} while the result 
evaluated by the $\chi^2$ minimization is shown in terms of
the full square at 0.8 GeV/c. Since the
predicted differential cross section at the diffractive minimum 
substantially depends on the antikaon momentum (by 
two orders of magnitude) we expect that the measurements should 
indicate a strong variation of $d\sigma/d\Omega$ even 
within the systematical uncertainties. 

\subsection{Antikaon momenta below 400~MeV/c}
Fig.~\ref{elastic8a} indicates that in free space the real 
part of the forward scattering amplitude close to zero antikaon 
momenta is negative for $K^-p$ and positive for $K^-n$
interactions. By taking the free space values the
real part of the optical potential for isospin symmetric nuclei
amounts to $\simeq$--18~MeV at $p_K{=}0$. Thus the real part
of $U_N$ is attractive, but small, which contradicts the 
experimental data on kaonic atoms where a depth of the $K^-$
potential at $\rho_0$ of 
--180~MeV~\cite{Gal1,Gal2} has been extracted. This descrepancy 
was resolved
in the literature by considering the in-medium 
modification~\cite{Koch,Waas1,Ohnishi,Lutz}  of the
$\Lambda(1405)$ resonance which couples to the $K^-p$
channel and yields a negative $f(0)$ in free space
at low antikaon momenta. The dominant effect, which modifies  the 
$\Lambda(1405)$ dynamics in nuclear matter, stems from the
Pauli blocking of the intermediate nucleon states which
causes a dissolution of this resonance~\cite{Brown3,Waas2}
and an in-medium modification of the $K^-p$ forward
scattering amplitude.
For the most recent review of the problem as well as the current 
status on the analysis  of kaonic atoms the reader is refered to
Ref.~\cite{Friedman}.

\begin{figure}[h]
\vspace{-4mm}
\psfig{file=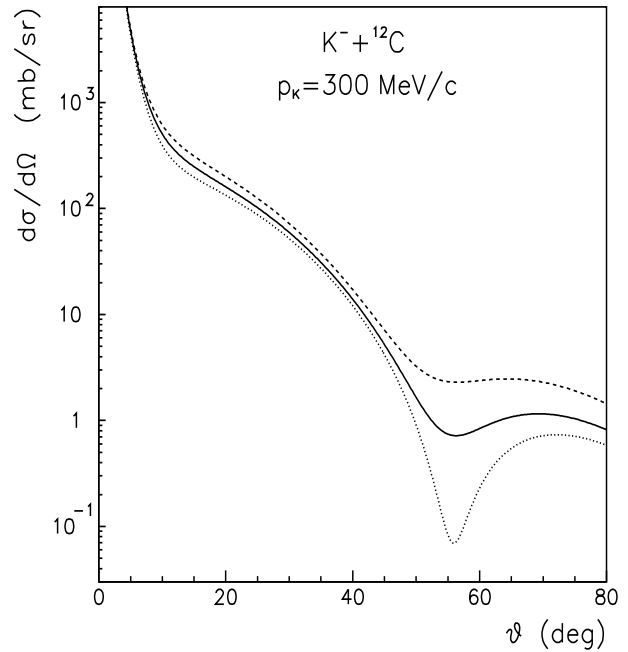,width=9cm,height=10.2cm}
\vspace{-2mm}
\caption[]{\label{elastic10}The differential cross section 
for $K^-C$ elastic scattering at $p_K$=300~MeV/c.
The solid line shows the calculation with $\sigma$ and $\alpha$
taken from free space, the dotted line shows the result for
$\alpha$=0, respectively. The dashed line represents a calculation 
where the contribution from the $\Lambda$(1405) resonance to the 
total real part of $f(0)$ is discarded.}
\end{figure}

The contribution from the $\Lambda(1405)$ resonance to the total
real part of the $K^-p$ forward scattering amplitude is
sizeable up to antikaon momenta of 
$\simeq$400~MeV/c~\cite{Sibirtsev1}. Thus 
the in-medium modification of $f(0)$ might also be studied
by $K^-$ elastic scattering from nuclei at low antikaon 
momenta $p_K{<}$400~MeV/c. 
However, the crucial question here is the validity of the Glauber 
theory at low energies. The limits of its
applicability as well as the analysis
of data on low energy antiproton elastic and inelastic 
scatterings from nuclei are given in Ref.~\cite{Dalkarov1}.
It follows that  Glauber theory at low
energies may serve only as a leading consideration.

Fig.~\ref{elastic10} shows the differential cross section for
$K^-$ elastic scattering from $^{12}C$ at an antikaon momentum
of 300~MeV/c. The solid line indicates a calculation with
cross section $\sigma$ and ratio $\alpha$ taken from the
free space $K^-N$ interactions while the dotted line
shows the result obtained with the same $\sigma$ but 
for $\alpha$=0.
 
The dashed line in Fig.~\ref{elastic10} shows the calculation 
with the modified real part of the forward $K^-p$ scattering 
amplitude by discarding the contribution from the  
$\Lambda$(1405) resonance to $Ref(0)$. In the latter case we obtain
$\alpha$=0.82 after averaging over protons and neutrons.

Fig.~\ref{elastic10} indicates that the difference between the 
calculations with different ratios $\alpha$ is sizeable. Although
our predictions at $p_K$=300~MeV/c may serve only as a leading 
extrapolation, we expect that the effect of the in-medium modification
of the forward scattering amplitude can actually be measured by elastic
$K^-A$ scattering.

\section{Conclusions}
In this work we have considered the possibility to extract the antikaon
potential from differential data on elastic $K^-A$ scattering
within the Glauber theory and analyzed the experimental 
results~\cite{Marlow} for $K^-$ elastic scattering from $^{12}C$ 
and $^{40}Ca$ at an antikaon momentum of 800~MeV/c. 
We have found that the total 
$K^-N$ cross section extracted from the data is close
to the value given in free space $K^-p$ and $K^-n$ interactions when
averaged over the number of protons and neutrons in the target.
On the other hand, the ratio $\alpha$ of the real to imaginary 
part of the forward scattering amplitude $f(0)$ - as evaluated from 
the data - differs from the ratio $\alpha$ in free space. Since 
the imaginary part of $f(0)$ seems not to be modified
this difference indicates an in-medium modification
of the real part of $f(0)$. However, the confidence level
of this result (within the $\chi^2$ method) is below
10\% which leaves it as an indication and does not allow for
a final conclusion.
The parameters evaluated from $K^-$ scattering from  
$^{12}C$ and $^{40}Ca$ have been used for a reconstruction of 
the antikaon potential at $p_K$=800~MeV/c, which is found to be attractive
in line with the analysis from Kiselev \cite{Kiselev} at 1.2 GeV/c.

We have investigated the target nuclear densities that can be 
tested by $K^-A$ elastic scattering and found them to range from 
0 to $\rho$=0.1~fm$^{-3}$. To extrapolate the depth of the
$K^-$ potential at $\rho{\simeq}$0.17~fm$^{-3}$ we have used nuclear 
density profiles with parameters evaluated from  
electron-nucleus scattering.

\begin{figure}[h]
\vspace{-6mm}
\psfig{file=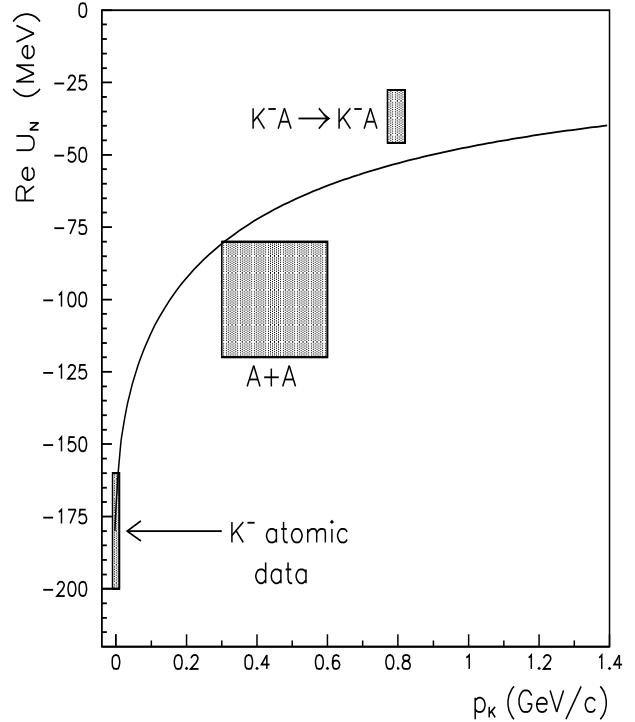,width=9cm,height=11.cm}
\vspace{-2mm}
\caption[]{\label{elastic11}The real part of the optical antikaon
potential at baryon density $\rho$=0.17~fm$^{-3}$ as a function of 
the $K^-$-meson momentum. The potential $Re\,U_N$ from the $K^-$ 
atomic data is taken from the recent analysis of 
Ref.~\cite{Friedman}. The  potential denoted by $A+A$ has been 
extracted from $K^-$ data on heavy-ion 
collisions~\protect\cite{Schroter,Senger1,Barth,Laue,Senger2} 
and is taken from 
Refs.~\protect\cite{Li1,Cassing1,Li2,Bratkovskaya,Cassing2}.
The potential evaluated from $K^-A{\to}K^-A$ elastic scattering
at 0.8~GeV/c 
is the result of the present study. The solid line shows our 
calculation from Ref.~\cite{Sibirtsev1}.}
\end{figure}

Furthermore, we collect the information on the real part of the antikaon 
potential from all available sources for different antikaon
momenta and show the results in Fig.~\ref{elastic11} 
extrapolated to normal nuclear density 
$\rho$=0.17~fm$^{-3}$. The potential $Re\,U_N{=}180{\pm}20$~MeV 
at $p_K$=0 stems from the data on kaonic atoms~\cite{Gal1,Gal2,Friedman}
with the most recent analysis performed in Ref.~\cite{Friedman}.
The analyis of the experimental 
results~\cite{Schroter,Senger1,Barth,Laue,Senger2} on 
$K^-$-meson production from heavy-ion collisions is taken from
Refs.~\cite{Li1,Cassing1,Li2,Bratkovskaya,Cassing2}
and indicates $Re\,U_N$=--80$\div$120~MeV at antikaon
momenta $300{\le}p_K{\le}600$~MeV/c. 
Fig.~\ref{elastic11} also shows the potential 
evaluated in this work from the
data on $K^-C$ and $K^-Ca$ elastic scattering; the indicated 
errorbar is due to a standard deviation corresponding to a 
5\% confidence level.
Moreover, the solid line in Fig.~\ref{elastic11} shows our 
result from Ref.~\cite{Sibirtsev1} calculated by discarding 
the contribution from $\Lambda$(1405) and $\Sigma$(1385)
resonances at nuclear matter density. Note that the experimental results 
as well as the calculations indicate an attractive $K^-$ potential up to
antikaon momenta of $\simeq$1.3~GeV/c. 

We, furthermore, have discussed future perspectives in 
$K^-A$ elastic scatterings
and the  investigation of the antikaon 
potential or modification of the scattering amplitude $f(0)$. 
It is pointed out that $K^-A$ elastic scattering at low antikaon 
momenta  may provide an additional (obvious) way to 
reconstruct  the antikaon potential as an alternative  to 
heavy-ion and  proton-
nucleus~\cite{Sibirtsev1,Kirchner,Sistemich,Schult,Sibirtsev3,Buescher} 
studies. Furthermore, measurements of the elastic $K^-$ 
scattering from nuclei at $p_K{\simeq}1.2$~GeV/c
provide an effective tool to investigate, if the real part of the 
$K^-n$ forward scattering amplitudy changes its sign around this
momentum, and to learn more about the interaction between the
antikaon and neutron.

\end{document}